\documentclass[12pt,twoside]{article}
\usepackage{amsmath} 
\usepackage{indentfirst}
\usepackage{dcolumn}
\usepackage{indentfirst}
\usepackage{graphicx}

\begin{document}

\title{A chiral symmetry breaking AdS / QCD model with scalar interactions}
\author{Alfredo Vega and Ivan Schmidt\\
\small Departamento de F\'\i sica y Centro Cient\'\i fico Tecnol\'ogico de Valpara\'\i so,\\
\small Universidad T\'ecnica Federico Santa Mar\'\i a,\\
\small Casilla 110-V, Valpara\'\i so, Chile}

%\date{\today}

\maketitle

\begin{abstract}

%Estudiamos modelos de tipo AdS / QCD con un potencial c\'ubico y cu\'artico para el sector escalar en el bulk considerando que la masa de los modos escalares depende de la coordenada hologr\'afica. En este tipo de modelos, a diferencia de lo que sucede al considerar masas variables sin estos t\'eminos adicionales, vemos que en general es posible obtener buenos espectros mes\'onicos considerando masas de quarks de corrientes.
We consider chiral symmetry breaking AdS / QCD models with cubic and
quartic potentials in the scalar sector and a z dependent mass for
scalars in the bulk. In these models it is possible to get good
mesonic spectra using current quark masses.
\end{abstract}

\date{\today}

\maketitle

%\section{Introducci\'on}
\section{Introduction}

%La llamada correspondencia Gauge / Gravedad actualmente se aplica en varias \'areas de la f\'isica te\'orica, entre las que destacan sus usos en  QCD, entregando una nueva metodolog\'ia para estudiar propiedades de la interacci\'on fuerte a bajas energ\'ia. Dentro del rango de propiedades hadr\'onicas que pueden ser estudiadas usando modelos AdS / QCD, encontramos modelos que incorporan expl\'icitamente los efectos del rompimiento de simetr\'ia quiral tanto de forma espont\'anea como explicitamente en el lagrangiano. Los primeros de estos modelos eran del tipo hard wall [~~], cuyo principal inconveniente es que el espectro hadr\'onico que se obtiene con ellos no reproducen espectros tipo Regge. Esta situaci\'on puede remediarse en el sector mes\'onico introduciendo un dilat\'on dependiente de z [~~], el que no funciona con bariones [~~], lo que requiere buscar alternativas complementarias al dilaton, como podr\'ia ser el considerar deformaciones a la m\'etrica [~~] y/o considerar masas variables [~~] para los modos que se propagan en el bulk describiendo a los hadrones, con lo que es posible obtener masas bari\'onicas con comportamiento Regge. Adicionalmente, el uso de masas dependientes de z en modelos hologr\'aficos resulta interesante pues podr\'ia tomar en cuenta en efecto de dimesiones an\'omalas de los operadores descritos por los modos AdS que se propagan en el bulk.

The gauge / gravity duality is currently applied to several areas in
theoretical physics. In the QCD case it provides a new methodology
to study strong interaction properties at low energy. Within the
range of hadronic properties that can be studied using AdS / QCD
models, we find models that consider effects of chiral symmetry
breaking in the Lagrangian. At the beginning these models were of
the hard wall type \cite{Erlich:2005qh,Da Rold:2005zs}, and their
main problem is that they cannot describe correctly the hadronic
spectra. This situation can be improved in the mesonic sector by
introducing a dilaton field that depends on the holographical
coordinate z \cite{Karch:2006pv, Colangelo:2008us}, although this
procedure does not work for baryons \cite{Kirsch:2006he,
Vega:2008te}. Here it is necessary to look for complementary
alternatives to the dilaton, such  as using a warp factor in the AdS
metric \cite{Forkel:2007cm},  and / or to consider a z dependent
mass for modes propagating in the bulk, which describe hadrons
\cite{Vega:2008te, Abidin:2009hr, Forkel:2008un, Forkel:2010gu}.
Additionally, variable masses in holographical models are
interesting to consider because in this way we can take into account
effects of anomalous dimensions of operators described by AdS modes
\cite{Cherman:2008eh, Vega:2008te}.

%La idea de usar masas variables en modelos AdS / QCD que consideren rompimiento de simetr\'ia quiral ha sido empleada en un modelo de tipo soft wall que considera rompimiento de simetr\'ia quiral [~~], en un espacio $AdS_{5}$ con dilat\'on cuadr\'atico sin interacciones en el sector escalar. El siguiente paso, que da origen a este trabajo, es estudiar el efecto de una masa variable en modelos que consideran interacciones en el sector escalar.

The use of variable mass in AdS / QCD models with chiral symmetry
breaking was considered in \cite{Vega:2010ne}, in a AdS spacetime
with quadratic dilaton, and without  interactions in the scalar
sector. The natural next step is to study models which include these
interactions; here we discuss the effect of a z
dependent mass in holographical models that consider cubic
\cite{Zhang:2010tk} and quartic \cite{Gherghetta:2009ac}
interactions in the scalar sector.

%En [~~] se ve que al usar masas variables en el modelo que no considera interacciones en el sector escalar la obtenci\'on de espectros mes\'onicos aceptables que consideren masas de corrientes para los quarks restringen el vev cuando $z \rightarrow \infty$ a un comportamiento lineal. Como veremos luego, la inclusi\'on de interacciones c\'ubicas [~~] o cu\'articas [~~] junto a la masa dependiente de z permite obtener modelos que reproducen buenos espectros mes\'onico usando masas de corrientes para los quarks tomando en cuenta cualquiera de los comportamientos posibles para el vev en el infinito.

In \cite{Vega:2010ne}, in order to get good mesonic spectra with
current quark masses in the model, it was necessary to consider a
vacuum expectation value (vev) with linear behavior when $z
\rightarrow \infty$. As we will see below, the inclusion of cubic
\cite{Zhang:2010tk} and quartic \cite{Gherghetta:2009ac}
interactions plus z dependent masses allows to get spectra with
Regge behavior, with current quark mass values, and without the
above restriction on the vev.

%\section{Modelo}
\section{Model}

%Consideramos un modelo soft wall AdS / QCD como el usado usando en [VegaSchmidt], al que adicionamos terminos c\'ubicos y cu\'artico en el potencial escalar del bulk. El modelo considera un background de tipo AdS en 5D definido por la m\'etrica
We consider a soft wall AdS / QCD model as in \cite{Vega:2010ne},
but we add cubic \cite{Zhang:2010tk} or quartic
\cite{Gherghetta:2009ac} potentials in the bulk scalar potential.
The 5d AdS background is defined by
\begin{equation}
 \label{Metrica}
 d s^{2} = \frac{R^{2}}{z^{2}} (\eta_{\mu \nu} dx^{\mu} dx^{\nu} + dz^{2}),
\end{equation}
%donde R es el radio de curvatura AdS, la m\'etrica de Minkowski $\eta_{\mu \nu} = diag (-1, +1, +1, +1)$ y z corresponde a la coordenada hologr\'afica definida en el rango $0 \leq z < \infty$. En este trabajo consideraremos un dilat\'on cuadr\'atico usual
where R is the AdS radius, the Minkowski metric is $\eta_{\mu \nu} =
diag (-1, +1, +1, +1)$ and z is a holographical coordinate defined
in the range $0 \leq z < \infty$. In this paper we consider a usual
quadratic dilaton $\phi (z) = \lambda^{2} z^{2}$. To describe chiral symmetry breaking in the mesonic sector and in
the 5d AdS side, the action contains $SU(2)_{L} \times SU(2)_{R}$
gauge fields and a scalar field X, and it is given by
%\begin{widetext}
\begin{equation}
%\label{Accion}
 S_{5} = - \int d^{5}x \sqrt{-g} e^{-\phi(z)} Tr \biggl[|DX|^{2} + m_{X}^{2} (z) |X|^{2} - \kappa |X|^{n} \nonumber
\end{equation}
\begin{equation}
\label{Accion}
 + \frac{1}{4 g_{5}^{2}} (F_{L}^{2} + F_{R}^{2})\biggr].
\end{equation}
%Esta acci\'on muestra de forma expl\'icita que la masa a considerar para los modos escalares es dependiente de z, ademas de incluir un potencial en el sector escalar, que en este trabajo ser\'a c\'ubico o cu\'atico seg\'un n sea 3 o 4.
This action shows explicitly that the scalar modes masses are z
dependent, besides including cubic or quartic bulk scalar potentials
for n equal 3 or 4.

%Aqu\'i $g_{5}^{2} = \frac{12 \pi^{2}}{N_{c}}$, con $N_{c}$ el n\'umero de colores, y los campos $F_{L,R}$ estan definidos por
Here $g_{5}^{2} = \frac{12 \pi^{2}}{N_{c}}$, where $N_{c}$ is the
number of colors, and the fields $F_{L,R}$ are defined by
\begin{equation}
 F_{L,R}^{MN} = \partial^{M} A_{L,R}^{N} - \partial^{N} A_{L,R}^{M} - i [A_{L,R}^{M},A_{L,R}^{N}], \nonumber
\end{equation}
%aqu\'i $A_{L,R}^{MN} = A_{L,R}^{MN} t^{a}$, $Tr[t^{a} t^{b}] = \frac{1}{2} \delta^{ab}$, y la derivada covariante es
here $A_{L,R}^{MN} = A_{L,R}^{MN} t^{a}$, $Tr[t^{a} t^{b}] = \frac{1}{2} \delta^{ab}$, and the covariant derivative is
\begin{equation}
 D^{M}X = \partial^{M} X - i A_{L}^{M} X + i X A_{R}^{M}. \nonumber
\end{equation}

\begin{table}
\begin{center}
\caption{Field content and dictionary of the model.}
\begin{tabular}{ c c c | c c c | c c c | c c c | c c c }
  \hline
  \hline
  & 4D : \textit{O}(x) & & & 5D : $\Phi(x,z)$ & & & p & & & $\Delta$ & & & $m_{5}^{2} R^{2}$ & \\
  \hline
  & $\overline{q}_{L} \gamma^{\mu} t^{a} q_{L}$ & & & $A_{L \mu}^{a}$ & & & 1 & & & 3 & & & 0 & \\
  & $\overline{q}_{R} \gamma^{\mu} t^{a} q_{R}$ & & & $A_{R \mu}^{a}$ & & & 1 & & & 3 & & & 0 & \\
  & $\overline{q}_{R}^{\alpha} q_{L}^{\beta}$ & & & $\frac{1}{z} X$ & & & 0 & & & $3 + \delta$ & & & $m_{5}^{2}(z) R^{2}$ & \\
  \hline
  \hline
\end{tabular}
\end{center}
\end{table}

%El campo escalar X, que es dual al operador $q\overline{q}$, tiene un vev dado por
The scalar field X, which is dual to the operator $q\overline{q}$, has a vev given by $X_{0} = \frac{v(z)}{2}$, which produces chiral symmetry breaking.

%En la tabla I aparecen los campos incluidos en el modelo y su relaci\'on con los modos que se propagan en el bulk de acuerdo con el diccionario AdS / CFT, donde hacemos incapi\'e en que el operador $q\overline{q}$ posee una dimensi\'on an\'omala ($\delta$), la que se traduce en una masa dependiente de z para los modos duales a dicho operador, de acuerdo con \cite{Cherman:2008eh, Vega:2008te}.
In Table I we show the fields included in the model, and also their
relationship with modes propagating in the bulk,  according to the
AdS / CFT dictionary. Notice that the operator $q\overline{q}$  has
an anomalous dimension, which in turn produces a mass
that depends on z for the modes dual to this operator, in agreement
with \cite{Cherman:2008eh, Vega:2008te}.

%A partir de (\ref{Accion}), las ecuaciones que describen tanto al vev, como a mesones escalares cambian seg\'un se considere un potencial  c\'ubico o cu\'artico
Starting from (\ref{Accion}), the equations that describe the vev
and the scalar, change if we consider n equal 3 or 4.
%Para n igual a 3 tenemos
For $n=3$ we have
\begin{equation}
 - z^{2} \partial_{z}^{2} v(z) + z ( 3 + 2 \lambda^{2} z^{2} ) \partial_{z} v(z) + m_{X}^{2} (z) R^{2} v(z) \nonumber
\end{equation}
\begin{equation}
 \label{vev1}
 - \dfrac{3}{4} R^{2} \kappa v(z)^{2} = 0,
\end{equation}

\begin{equation}
 - \partial_{z}^{2} S_{n} (z) + \frac{3+2 \kappa^{2} z^{2}}{z} \partial_{z} S_{n} (z) + \frac{m_{X}^{2} (z) R^{2}}{z^{2}} S_{n} (z) \nonumber
\end{equation}
\begin{equation}
 \label{escalar1}
 - \frac{R^{2} \kappa v(z)}{z^{2}} S_{n} (z) = M_{S}^{2} S_{n}(z),
\end{equation}

%y cuando $n=4$ se tiene
and when $n=4$,
\begin{equation}
 - z^{2} \partial_{z}^{2} v(z) + z ( 3 + 2 \lambda^{2} z^{2} ) \partial_{z} v(z) + m_{X}^{2} (z) R^{2} v(z) \nonumber
\end{equation}
\begin{equation}
 \label{vev2}
 - \dfrac{1}{2} R^{2} \kappa v(z)^{3} = 0,
\end{equation}

\begin{equation}
 - \partial_{z}^{2} S_{n} (z) + \frac{3+2 \kappa^{2} z^{2}}{z} \partial_{z} S_{n} (z) + \frac{m_{X}^{2} (z) R^{2}}{z^{2}} S_{n} (z) \nonumber
\end{equation}
\begin{equation}
 \label{escalar2}
 - \frac{3 R^{2} \kappa v^{2}(z)}{2 z^{2}} S_{n} (z) = M_{S}^{2} S_{n}(z),
\end{equation}

%Por otro lado, para mesones vectores y vectores axiales se tiene que las ecuaciones son iguales en ambos casos
On the other hand, vector and axial vector equations are the same in
both cases
\begin{equation}
 \label{vector}
 - \partial_{z}^{2} V_{n} (z) + \biggl( \frac{1}{z} + 2 \lambda^{2} z \biggr) \partial_{z} V_{n} (z) = M_{V}^{2} V_{n}(z),
\end{equation}
\begin{equation}
 - \partial_{z}^{2} A_{n} (z) + \biggl( \frac{1}{z} + 2 \lambda^{2} z \biggr) \partial_{z} A_{n} (z) + \nonumber
\end{equation}
\begin{equation}
 \label{vector axial}
 \frac{R^{2} g_{5}^{2} v^{2} (z)}{z^{2}} A_{n} (z) = M_{A}^{2} A_{n}(z).
\end{equation}

%Para discutir la fenomenolo\'ia que se obtiene de estas ecuaciones, es necesario conocer la forma que tendr\'a $m_{X}^{2} (z)$, lo que hacemos de forma an\'aloga a lo realizado en \cite{Vega:2010ne}, la masa del modo escalar que se propaga en el bulk $m_{X}^{2} (z)$, se obtiene a partir de (\ref{vev1}) o (\ref{vev2}) suponiendo conocida la forma de $v(z)$, que corresponder\'a a una funci\'on capaz de reproducir un par de l\'imites conocidos.
To discuss the phenomenology of this model, it is necessary to know
the precise form of $m_{X}^{2} (z)$. This can be done as in Ref.
\cite{Vega:2010ne}. The mass for scalar modes in the bulk,
$m_{X}^{2} (z)$, is obtained starting from (\ref{vev1}) for the
cubic case and from (\ref{vev2}) in the quartic case, provided that
the function $v(z)$ is known.

\begin{table}[hb]
\begin{center}
\caption{Parameters used in the model.}
\begin{tabular}{ c c c | c c c | c c c | c c c }
  \hline
  \hline
  & ~~ & & & $c_{I}=0.4$ & & & $m_{q}=4.3 MeV$ & & & $R \kappa=-11$ & \\
  & Cubic & & & $c_{II}=28$ & & & $m_{q}=0.8 MeV$ & & & $R \kappa=-1$ & \\
  & ~~ & & & $c_{III}=3$ & & & $m_{q}=0.8 MeV$ & & & $R \kappa=-10$ & \\
  \hline
  & ~~ & & & $c_{I}=0.4$ & & & $m_{q}=4.3 MeV$ & & & $\kappa=-11$ & \\
  & Quartic & & & $c_{II}=28$ & & & $m_{q}=0.8 MeV$ & & & $\kappa=-0.02$ & \\
  & ~~ & & & $c_{III}=3$ & & & $m_{q}=0.8 MeV$ & & & $\kappa=-0.76$ & \\
  \hline
  & ~~ &&& ~~ &&& $\lambda=0.4 GeV$ &&& ~~ & \\
  \hline
  \hline
\end{tabular}
\end{center}
\end{table}

The behavior of $v(z)$ can be known in two limits.
%En primer lugar consideramos el l\'imite usual para $z \rightarrow 0$, seg\'un el cual
First we consider the usual limit $z \rightarrow 0$, according to which
\begin{equation}
 \label{vev Cero}
v(z \rightarrow 0) = \alpha z + \beta z^{3},
\end{equation}
%donde los coeficientes $\alpha$ y $\beta$ est\'an asociados con la masa de los quarks y el condensado quiral respectivamente.
where the $\alpha$ and $\beta$ coefficients are associated with the
quark mass and chiral condensate respectively.

%In \cite{Vega:2010ne} the authors discuss a model with z dependent mass in the scalar sector, in \cite{Zhang:2010tk} a model with cubic interaction was discussed, and \cite{Gherghetta:2009ac} presents a model with quartic interaction. Notice that the mass of the lightest scalar meson in \cite{Zhang:2010tk} is less than the mass of the pion, contradicting a well-established QCD theorem  \cite{Weingarten:1983uj, Witten:1983ut}.

\begin{table*}[ht]
\begin{center}
\caption{Scalar meson spectra in cubic case. We give experimental values, masses calculated in each model considered here, and compare them with a couple of holographical models. All masses are in MeV. Notice that the mass of the lightest scalar meson in \cite{Zhang:2010tk} is less than the mass of the pion, contradicting a well-established QCD theorem  \cite{Weingarten:1983uj, Witten:1983ut}.}
\scalebox{0.7}{\begin{tabular}{ c c c | c c c | c c c | c c c | c c c | c c c | c c c | c c c }
  \hline
  \hline
  & n & & & $f_{0} (Exp)$ & & & $f_{0}$   & & & $f_{0}$    & & & $f_{0}$   & & & $f_{0} (Ref.\cite{Vega:2010ne})$  & & & $f_{0} (Ref. \cite{Zhang:2010tk})$   & & & $f_{0} (Ref. \cite{Gherghetta:2009ac})$ \\
  &   & & &               & & & $m_{q} = 4.3$ & & & $m_{q} = 0.8$ & & & $m_{q} = 0.8$ & & & ~  & & & ~  & & &   \\
  &   & & &               & & & $c_{_{I}} = 0.4$ & & & $c_{_{II}} = 28$ & & & $c_{_{III}} = 3 $ & & & ~ & & &  ~  & & &  \\
  \hline
  & 0 & & & $550^{+250}_{-150}$ & & & 487 & & & 552 & & & 555 & & & 485 & & & 118 & & & 799 & \\
  & 1 & & & $980 \pm 10$ & & & 1193 & & & 916 & & & 986 & & & 903 & & & 953 & & & 1184 & \\
  & 2 & & & $1350 \pm 150$ & & & 1452 & & & 1213 & & & 1275 & & & 1208 & & & 1335 & & & 1466 & \\
  & 3 & & & $1505 \pm 6$ & & & 1665 & & & 1454 & & & 1509 & & & 1451 & & & 1627 & & & 1699 & \\
  & 4 & & & $1724 \pm 7$ & & & 1851 & & & 1661 & & & 1710 & & & 1659 & & & 1873 & & & 1903 & \\
  & 5 & & & $1992 \pm 16$ & & & 2019 & & & 1845 & & & 1890 & & & 1844 & & & 2089 & & & 2087 & \\
  & 6 & & & $2103 \pm 8$ & & & 2172 & & & 2012 & & & 2054 & & & 2012 & & & 2285 & & & 2257 &\\
  & 7 & & & $2314 \pm 25$ & & & 2316 & & & 2166 & & & 2205 & & & 2166 & & & 2465 & & & 2414 &\\
  \hline
  \hline
\end{tabular}}
\end{center}
\end{table*}

\begin{table*}[ht]
\begin{center}
\caption{As in Table III, but for scalar meson spectra in quartic case.}
\scalebox{0.7}{\begin{tabular}{ c c c | c c c | c c c | c c c | c c c | c c c | c c c | c c c }
  \hline
  \hline
  & n & & & $f_{0} (Exp)$ & & & $f_{0}$   & & & $f_{0}$    & & & $f_{0}$   & & & $f_{0} (Ref.\cite{Vega:2010ne})$  & & & $f_{0} (Ref.\cite{Zhang:2010tk})$   & & & $f_{0} (Ref.\cite{Gherghetta:2009ac})$ \\
  &   & & &               & & & $m_{q} = 4.3$ & & & $m_{q} = 0.8$ & & & $m_{q} = 0.8$ & & & ~  & & & ~  & & &   \\
  &   & & &               & & & $c_{_{I}} = 0.4$ & & & $c_{_{II}} = 28$ & & & $c_{_{III}} = 3$ & & & ~ & & & ~  & & &   \\
  \hline
  & 0 & & & $550^{+250}_{-150}$ & & & 548 & & & 595 & & & 553 & & & 485 & & & 118 & & & 799 & \\
  & 1 & & & $980 \pm 10$ & & & 1232 & & & 1146 & & & 1251 & & & 903 & & & 953 & & & 1184 & \\
  & 2 & & & $1350 \pm 150$ & & & 1477 & & & 1478 & & & 1675 & & & 1208 & & & 1335 & & & 1466 & \\
  & 3 & & & $1505 \pm 6$ & & & 1683 & & & 1661 & & & 2010 & & & 1451 & & & 1627 & & & 1699 & \\
  & 4 & & & $1724 \pm 7$ & & & 1865 & & & 1831 & & & 2296 & & & 1659 & & & 1873 & & & 1903 & \\
  & 5 & & & $1992 \pm 16$ & & & 2030 & & & 1989 & & & 2549 & & & 1844 & & & 2089 & & & 2087 & \\
  & 6 & & & $2103 \pm 8$ & & & 2182 & & & 2137 & & & 2779 & & & 2012 & & & 2285 & & & 2257 &\\
  & 7 & & & $2314 \pm 25$ & & & 2324 & & & 2276 & & & 2991 & & & 2166 & & & 2465 & & & 2414 &\\
  \hline
  \hline
\end{tabular}}
\end{center}
\end{table*}

\begin{table*}[ht]
\begin{center}
\caption{Axial vector mesons spectra. We give experimental values,
masses calculated in each model considered here, and compare them
with a couple of holographical models. All masses are in MeV.}
\scalebox{0.7}{\begin{tabular}{ c c c | c c c | c c c | c c c | c c c | c c c | c c c | c c c }
  \hline
  \hline
  & n & & & $a_{1} (Exp)$ & & & $a_{1}$ & & & $a_{1}$ & & & $a_{1}$ & & & $a_{1} (Ref.\cite{Vega:2010ne})$ & & & $a_{1} (Ref.\cite{Zhang:2010tk})$ & & &  $a_{1} (Ref.\cite{Gherghetta:2009ac})$ & \\
  &   & & &               & & & $m_{q} = 4.3$ & & & $m_{q} = 0.8$ & & & $m_{q} = 0.8$ & & & ~~ & & & ~~ & & &    \\
  &   & & &               & & & $c_{_{I}} = 0.4$ & & & $c_{_{II}} = 28$ & & & $c_{_{III}} = 3$ & & & ~~ & & & ~~  & & &    \\
  \hline
  & 0 & & & $1230 \pm 40$ & & & 1236 & & & 804 & & & 1778 & & & 811 & & & 997 & & & 1185 & \\
  & 1 & & & $1647 \pm 22$ & & & 1412 & & & 1135 & & & 2526 & & & 1133 & & & 1541 & & & 1591 & \\
  & 2 & & & $1930^{+39}_{-70}$ & & & 1563 & & & 1388 & & & 3099 & & & 1384 & & & 1934 & & & 1900 & \\
  & 3 & & & $2096 \pm 122$ & & & 1731 & & & 1602 & & & 3582 & & & 1601 & & & 2258 & & & 2101 & \\
  & 4 & & & $2270^{+55}_{-40}$ & & & 1896 & & & 1791 & & & 4006 & & & 1789 & & & 2540 & & & 2279 & \\
  \hline
  \hline
\end{tabular}}
\end{center}
\end{table*}

%El otro l\'imite considerado para $v(z)$, es $z \rightarrow \infty$. En esta caso exigimos que (\ref{vector axial}) entregue un espectro con comportamiento en el l\'imite de z grande. Como el potencial c\'ubico o cu\'artico no se acopla a la ecuaci\'on que describe a mesones vectores axiales, esta es igual en los casos que discutimos, y es igual a la que se obtiene cuando no se consideran potenciales escalares en el bulk, obteni\'endose las mismas expresiones que se hallaron en [VEGA SCHMIDT], donde se vio que $v(z \rightarrow \infty)$ puede ser contante, lineal o cuadr\'atica en z.
The other limit for $v(z)$ is when $z \rightarrow \infty$. Here we
consider the condition that we should get Regge spectra when z is
high in equation (\ref{vector axial}). In fact, equation
(\ref{vector axial}) does not change when $\kappa = 0$, which was
the case analyzed in \cite{Vega:2010ne}, so we can use the
expressions considered obtained there, where it was shown that $v(z
\rightarrow \infty)$ can be constant, linear or quadratic in z.
%Por esta raz\'on consideramos los ansatze usados en [VEGA SCHMIDT] que interpolan los l\'imites mencionados para $v(z)$.
For this reason we use the ans\"atze considered in
\cite{Vega:2010ne} to reproduce both limits for each possible z
behavior in the high z limit.
\begin{equation}
 \label{Ansatz vev I}
v_{_{I}} (z) = \frac{c_{_{I}}}{R} \arctan (A z + B z^{3}), ~~~~(Model~I)
\end{equation}
\begin{equation}
 \label{Ansatz vev II}
v_{_{II}} (z) = \frac{z}{R} (A + B \tanh (c_{_{II}} z^{2})), ~~~~(Model~II)
\end{equation}
\begin{equation}
 \label{Ansatz vev III}
v_{_{III}} (z) = \frac{A z + B z^{3}}{R \sqrt{1 + c_{_{III}}^{2} z^{2}}} . ~~~~(Model~III)
\end{equation}

%En relaci\'on a los par\'ametros del modelo, $\lambda$ usualmente se fija con datos de espectros hadr\'onicos. Hemos escogido $\lambda = 0.400 GeV$, con lo que se obtiene un buen acuerdo con las masas de mesones vectoriales.
In relation to the parameters of model, $\lambda$ is fixed using
data from the spectrum. We choose $\lambda = 0.400 GeV$, which
allows us to obtain correct mass values for vector mesons. The
remaining parameters can be fixed using (\ref{Ansatz vev I}),
(\ref{Ansatz vev II}) and (\ref{Ansatz vev III}). Comparing with the
value established in the AdS / CFT dictionary, with the notation
used in \cite{Gherghetta:2009ac}
\begin{equation}
 \label{vev UV}
v (z \rightarrow 0) = \frac{ m_{q} \zeta}{R} z + \frac{\sigma}{R \zeta} z^{3},
\end{equation}
% Ac\'a hemos usado el par\'ametro $\zeta$ introducido \cite{Cherman:2008eh}, que permite una correcta normalizaci\'on, y cuyo valor esta dado por $\zeta = \sqrt{3}/(2 \pi)$. Con esto, los par\'ametros A y B corresponden a
The parameter $\zeta$  was introduced in \cite{Cherman:2008eh} to
get the right normalization, and its value is   $\zeta = \sqrt{3}/(2
\pi)$.

%Para completar la descripci\'on del modelo, es necesario especificar los valores de $m_{q}$ y $\sigma$, pero debido a que ambas se encuentran relacionadas por medio de la relaci\'on GOR $m^{2}_{\pi} f^{2}_{\pi} = 2 m_{q} \sigma$, tan solo es necesario fijar una de ellas. En esta caso, consideramos $m_{\pi} = 139.6 MeV$ y $f_{\pi} = 92.4$ fijamos la masa de los quarks usando
In order to finish the model description, it is necessary to specify
the values for $m_{q}$ and $\sigma$, which are related by the Gell-Mann-Oakes-Renner relation
$m^{2}_{\pi} f^{2}_{\pi} = 2 m_{q} \sigma$, and therefore we need to
fix only one of them. In this case we use $m_{\pi} = 140~ MeV$ and
$f_{\pi} = 92~ MeV$, and we fix the quark mass using
\begin{equation}
 \label{Constante Decaimiento}
f^{2}_{\pi} = - \frac{1}{g^{2}_{5}} \lim_{\epsilon \rightarrow 0} \frac{\partial_{z} A_{0} (0,z)}{z} |_{z = \epsilon},
\end{equation}
%donde $A_{0} (0,z)$ es soluci\'on de (\ref{vector axial}), con $M^{2}_{A} = 0$, y las condiciones de borde usadas son $A_{0}(0,0)=1$ y $\partial_{z} A_{0} (0,z \rightarrow \infty) = 0$.
where $A_{0} (0,z)$ is solution of (\ref{vector axial}), with
$M^{2}_{A} = 0$, and the boundary conditions used are $A_{0}(0,0)=1$
and $\partial_{z} A_{0} (0,z \rightarrow \infty) = 0$.

%Si se observa (\ref{vector axial}), la ecuaci\'on para $A_{0} (0,z)$ posee un t\'ermino que depende de $m_{q}$, asi es que el uso de (\ref{Constante Decaimiento}) entrega un $f_{\pi} (m_{q})$, el que aparece graficado en FIG XXX, y muestra que por cada valor de $\Omega$ considerado, hay dos valores posibles para la masa de los quarks, que arrojan una constante de decaimiento para el pion igual a 92 MeV. Para $\Omega = 0.1$ encontramos $m_{q} = 2.8 MeV$ y $m_{q} = 74.3 MeV$; para $\Omega = 0.5$ hallamos $m_{q} = 7.9 MeV$ y $m_{q} = 72.3 MeV$ y cuando usamos $\Omega = 2$ se obtiene $m_{q} = 7.3 MeV$ y $m_{q} = 73.1 MeV$.
As can be observed in (\ref{vector axial}), the $A_{0} (0,z)$
equation as a term that depends on $m_{q}$, so using (\ref{Constante
Decaimiento}) we get $f_{\pi} (m_{q})$. In general we obtain two
possible quark masses in each model, and one of them can be
considered as a current mass.

\begin{table}[h t]
\begin{center}
\caption{Vector mesons spectra in MeV.}
\begin{tabular}{ c c c | c c c | c c c | c c c | c c c }
  \hline
  \hline
  & n & & & $\rho (Exp)$ & & & $\rho (Model)$  & & & $\rho (Ref.\cite{Zhang:2010tk})$ & & & $\rho (Ref.\cite{Gherghetta:2009ac})$ & \\
  \hline
  & 0 & & & $775.5 \pm 1$ & & & 800 & & & 759 & & & 475 & \\
  & 1 & & & $1282 \pm 37$ & & & 1131 & & & 1202 & & & 1129 & \\
  & 2 & & & $1465 \pm 25$ & & & 1386 & & & 1519 & & & 1529 & \\
  & 3 & & & $1720 \pm 20$ & & & 1600 & & & 1779 & & & 1674 & \\
  & 4 & & & $1909 \pm 30$ & & & 1789 & & & 2005 & & & 1884 & \\
  & 5 & & & $2149 \pm 17$ & & & 1960 & & & 2207 & & & 2072 & \\
  & 6 & & & $2265 \pm 40$ & & & 2117 & & & 2393 & & & 2243 & \\
  \hline
  \hline
\end{tabular}
\end{center}
\end{table}

%\section{Espectro mes\'onico}
\section{Mesonic spectrum}

%Los par\'ametros del modelo han sido fijados de modo an\'alogo a lo hecho en \cite{Vega:2010ne}, con lo que podemos calcular la masa de algunos mesones con el modelo, las que corresponden a los valores propios de (\ref{escalar}), (\ref{vector}) y (\ref{vector axial}). De estas tres ecuaciones, solamente (\ref{vector}) puede ser resuelta anal\'iticamente, por lo que como estrategia general optamos por transformarlas en ecuaciones de Schr\"odinger, para luego resolver num\'ericamente las ecuaciones que no poseen soluciones exactas, usando el programa de MATHEMATICA llamado schroedinger.nb \cite{Lucha:1998xc}, que ha sido adaptado para poder trabajar con los potenciales de las ecuaciones de Schr\"odinger que obtenemos. A partir de las ecuaciones (\ref{escalar1}), (\ref{escalar2}), (\ref{vector}), (\ref{vector axial}) se obtienen scuaciones tipo Schr\"odinger cuyos potenciales resultan ser.
The model parameters are summarized in Table II, and are fixed as in
\cite{Vega:2010ne}. We can then calculate meson masses, which
correspond to eigenvalues in the equations (\ref{escalar1}) or
(\ref{escalar2}), (\ref{vector}) and (\ref{vector axial}). In this
set of equations, only (\ref{vector}) can be solved analytically.
For this reason we prefer to change equations (\ref{escalar1}),
(\ref{escalar2}) and (\ref{vector axial}) into Schr\"odinger
like ones, and solve them numerically using a MATHEMATICA code
\cite{Lucha:1998xc}, which was adapted to our potentials.

For scalars, depending on wether we consider cubic or quartic
interactions, we have
\begin{equation}
 \label{Potencial Escalar Cubico}
V_{Sc} (z) = 2 \lambda^{2} + \lambda^{4} z^{2} + \frac{15}{4 z^{2}} + \frac{m_{X}^{2} (z) R^{2}}{z^{2}} - \dfrac{R^2 \kappa v(z)}{z^{2}}.
\end{equation}
\begin{equation}
 \label{Potencial Escalar Cuartico}
V_{Sq} (z) = 2 \lambda^{2} + \lambda^{4} z^{2} + \frac{15}{4 z^{2}} + \frac{m_{X}^{2} (z) R^{2}}{z^{2}}  - \dfrac{6 R^2 \kappa (v(z))^{2}}{z^{2}}.
\end{equation}

%Para mesones vectoriales se tiene el mismo potencial en ambos casos
For vector mesons, in all cases we get the same potential
\begin{equation}
 \label{PotencialVector}
V_{V} (z) = \frac{3}{4 z^{2}} + \lambda^{4} z^{2},
\end{equation}
%y es extraer un espectro exacto, que es de la forma
and in this case it is possible to get an exact spectrum,
\begin{equation}
 \label{Espectro Vectores}
M_{V}^{2} = 4 \lambda^{2} (n + 1).
\end{equation}

%Finalmente, para el caso de mesones vectores axiales el potencial es
Finally, for axial vector mesons the potential is
\begin{equation}
 \label{PotencialAxial 2.0}
V_{A} (z) = \frac{3}{4 z^{2}} + \lambda^{4} z^{2} + \frac{R^{2} g_{5}^{2} v^{2} (z)}{z^{2}}.
\end{equation}

%A diferencia de lo que sucede para mesones vectoriales, donde la soluci\'on es exacta, en los restantes casos el espectro se obtiene al resolver numericamente las ecuaciones asociadas. Las tablas XXX, YYY, ZZZ y QQQ son un resumen de las masas que se obtienen con el modelo.
The mesonic mass spectra that
we get are summarized in Tables III to VI, where we additionally
include masses calculated in other holographical models.

%\section{Conclusiones}
\section{Conclusions}

This paper is an extension of ideas discussed in a previous work \cite{Vega:2010ne}, taking into account cubic or quartic interactions in the scalar sector of the Lagrangian. Some specific equations are modified in relation to \cite{Vega:2010ne}, such as the vev equation, which produces a different $m_{x}(z)$. Changes in the $S_{n}$ equation, considering both cubic or quartic additional interactions, provide better spectra, allowing us to get good mesonic masses with current quark mass values, for every v(z) used.

The model used is of the Bottom-Up type, and the mesonic spectra is satisfactory, with an easy hadron identification, in contrast to what happens in models such as Sakai-Sugimoto \cite{Sakai:2004cn, Sakai:2005yt}, the best known Top-Down model, where there exist modes that cannot be identified with mesons in QCD.

Additionally we like to highlight that although the model was used in this paper to describe mesons, it is possible to apply it to study exotics \cite{Vega:2008af, Vega:2008te, Forkel:2010gu} (this application is postponed to future work), since the mass of scalar modes that describe hadrons in the bulk depends for these modes on the conformal dimension $\Delta$ according to $m_{5}^{2}R^{2}=\Delta(\Delta-4)$, and the dictionary tells us that the conformal dimension is related to operator dimensions, which looks like $\Delta_{0}+\delta$, where for states with angular momentum equal to zero $\Delta_{0}$ depends on the number and kind of hadrons constituents, and $\delta$ is the anomalous dimension. This opens up the possibility to consider z dependent masses of some AdS modes in this kind of models \cite{Cherman:2008eh, Vega:2010ne, Vega:2008te}. Here we have only considered mesons, but the possibility to consider exotics implies the use of the right $\Delta_{0}$ as in \cite{Vega:2008af, Vega:2008te, Forkel:2010gu}, and in particular this could allow us to differentiate between the lightest meson and a tetraquark from a holographic perspective, as was studied in detail in \cite{Forkel:2010gu} for an AdS / QCD model that considers chiral symmetry breaking.

The results obtained in this paper reinforce the idea that variable masses can be considered as a complementary alternative in holographic models. 

%\begin{Agradecimientos}
%\begin{acknowledgments}
\subsection*{Acknowledgments}

Work supported by Fondecyt (Chile) under Grants No. 3100028 and 1100287.

%\end{acknowledgments}

\end{document}